\documentclass[a4paper,11pt]{article}

\usepackage{jcappub} 

\usepackage{subfig}
\usepackage{multirow}

\title{Phase transition in compact stars: \\  nucleation mechanism and $\gamma$-ray bursts revisited}

\author{Kauan D. Marquez}
\author{and D\'ebora P. Menezes}
\affiliation{Departamento de Fisica, CFM - Universidade Federal de Santa Catarina;  C.P. 476, CEP 88.040-900, Florian\'opolis, SC, Brasil}

\emailAdd{kmarkez@hotmail.com}
\emailAdd{debora.p.m@ufsc.br}

\abstract{We have revisited the nucleation process based on the Lifshitz-Kagan theory, which is the underlying mechanism of   conversion of a pulsar constituted of hadronic matter to a quark star.  We have selected appropriate models that have been tested against experimental and observational constraints to restrict the model arbitrariness present in previous investigations. The phase transition pressures and chemical potentials have been identified and afterwards, the tunneling probabilities and the nucleation time were computed. The critical  pressures for which the half life of the metastable hadronic phase is one year were obtained. Even with the restrictions imposed to the selection of models, the results remained model dependent, but we found that the tunneling that makes possible the appearance of stable matter requires an overpressure that is practically independent of the quark matter bag constant. Finally, we have confirmed that the nucleation process can be one of the causes of gamma-ray bursts.}

\arxivnumber{1709.07040}

\begin{document}
\maketitle
\flushbottom

\section{Introduction}
\label{sec:intro}

Since its discovery in the late 60's \cite{bell}, the true nature of the ultra-dense compact objects known as {\it pulsars} remains, to some extent, undefined. This happens because of the unsolvability of the fundamental quantum field theory of the strong force, named quantum chromodynamics (QCD). Hence, a complete analytical description of the extremely dense matter in strongly interacting regime, as the one assumed to occur in the interior of this kind of object \cite{greiner1} is still not possible. 
As a consequence of this, both quali and quantitative descriptions of QCD matter depend on relativistic effective models \cite{schmitt}, implying that the internal structure and even the matter composition of such dense objects rely heavily on the equations of state (EoS) for hadronic matter derived from these models. 

The consensual, and rather simplistic, picture of pulsars considers them as neutron stars made of a homogeneous fluid of neutron-rich nuclear matter. Subsequent developments suggest that these objects could contain other exotic phases, such as heavier baryons or Bose-Einstein condensates together with the neutron matter, or even deconfined quark matter \cite{Glen}. Some results derived from models and observational data establish the mass of these compact stellar objects in the range $1.4<M/M_\odot<2.2$, its central density in the range $4<\rho/\rho_0<8$ and the radius of the order of $10$ km, with $M_\odot$ and $\rho_0$ standing for the solar mass and the nuclear saturation density, respectively \cite{astrolab,pmass}.

The Bodmer-Witten hypothesis \cite{ito,bw1,bw2,bw3} affirms that, although the totality of the physical terrestrial experiences attest that the fundamental state of baryonic matter presents confined quarks, it is not theoretically possible to confirm that as the real ground state of matter instead of a long-lasting metastable state. According to this idea, deconfined three-quark baryonic matter, named strange matter, might be energetically favored as compared with the two-quark ordinary hadronic matter because the inclusion of the $s$ quark in the ordinary $u$-$d$ matter represents a new freedom degree for the Fermi seas of the particles, lowering the total binding energy of the system \cite{qsweber}. {Moreover, the QCD phase diagram depicts a phase transition between certain density and  temperature domains. On one side, lies ordinary hadronic matter and on the other side the so-called quark-gluon plasma (QGP). According to lattice QCD simulations \cite{qgp1,qgp2,qgpl}, the transition between hadronic matter and the QGP is a crossover but, according to effective models, it is a first order phase transition. These two contradictory pictures can only be justified if a critical end point exists somewhere in the intersection of the transition curve coming from the lattice QCD domain, i.e., low chemical potential and high temperature, and the one obtained from effective models at high chemical potential and low temperature. We are next interested only in the latter, in which these variables are in the typical range expected for compact stars.}

In fact, the existence of stars made of deconfined quark matter was proposed by several researchers by the time of the assumption of quarks as the real fundamental particles of hadronic matter, see refs. \cite{iv1,iv2}, and still are theme of theoretical and observational investigation. Currently, there is no consensus on the existence of this type of object, although there are many evidences of their possible existence \cite{qsweber,qsexp1,qsexp2,qsexp3}. Also, even after a possible unambiguous observation of a strange star, it will be necessary to determine in which conditions pulsars might be identified as pure hadronic stars (neutron or hyperon stars, in the loose definition) or as deconfined quark stars (strange stars). Still, according to the Bodmer-Witten hypothesis, the hadronic matter in a neutron star interiors can be in an energetically unfavorable metastable state, which would allow its conversion into a strange star. 

Gamma ray bursts (GRBs) are cosmic high energy events known as the brightest electromagnetic events occurring in the universe. They can be distinguished mainly by their duration and released energies in long (LGRBs) and soft (SGRBs) gamma ray bursts \cite{grbdata1}. The total energy released in the first few hundred seconds by LGRBs is of the order of $10^{53}$ erg, which is about two orders greater than released in SGRBs \cite{grbdata2}. Due its extraordinary degree of diversity in terms of duration, luminosity, emission profile and spectra, almost completely unconstrained in terms of observational variables, the phenomenon (or, more plausibly, phenomena) that originates GRBs remains undefined.
Among many other possible progenitors, it was proposed that GRBs might be a manifestation of a phase transition inside compact stars, more precisely of the energy released in the conversion of a metastable hadronic star into a strange star \cite{cgrb1,cgrb2,debgrb}. 

Here we go further in the investigation of this possibility, extending the pure energetic approach of ref. \cite{debgrb} to account also for the conditions needed by the phase conversion, the lifetime of the metastable star and the phase transition mechanism. We restrict our investigation to relativistic models shown to describe presently accepted compact star properties, which satisfy nuclear matter and observational constraints \cite{const1,const2}. As far as the quark models are concerned, the MIT bag model is used \cite{mit}, but restricted to bag values that satisfy the stability window \cite{stab}. 

In section \ref{sec2} of the present work we expose the relativistic effective models used to describe the hadronic and quark phases, together with the application of the EoS to the construction of compact stars. In section \ref{sec3} we review the the phase transition dynamics, summarizing the formalism of the process believed to occur in this case, called nucleation, and then applying it to the compact star picture. Some astrophysical consequences of our results are then discussed.  In section \ref{sec4} the conclusions are drawn.

\section{Effective models of dense matter \label{sec2}}

In this section we give a brief summary of the effective models used in the calculations to describe dense matter in the QCD regime. Only the zero temperature regime is considered because cold compact stars are the ones that result in the evolutionary stage after the Urca process takes place \cite{urca}.
Relativistic mean filed approximation (RMF) is used in the derivation of the equations of state.

\subsection{Hadronic phase}

The relativistic effective model used here to describe the hadronic matter is a rather generalized version of the quantum hadrodynamics (QHD) \cite{qhd1,qhd2}, where the strong interaction is emulated by the exchange of massive mesons through Yukawa-type potentials, known as nonlinear Walecka model (NLWM). In this model, the interaction between baryons is mediated by the scalar mesons $\sigma$, scalar also in terms of isospin, and $\delta$, isovectorial, and by the vector mesons $\omega$, isoescalar, and $\rho$, vector with respect to both spin and isospin. 

The Lagrangian density of the NLWM, also known as Boguta-Bodmer model, for matter with hyperons, reads \cite{const1,bogbod,walecka},
\begin{eqnarray}
  \mathcal{L}_{\text{\tiny NLWM}} ={}& \sum_{B} \bar{\psi }_{B}\left [ \gamma ^{\mu }\left ( i\partial ^{\mu }-g_{\omega B}\omega _{\mu }-\frac{1}{2}g_{\rho B}\vec{\tau }\cdot \vec{\rho }_{\mu } \right )-\left ( M_{B}-g_{\sigma B}\sigma -g_{\delta B}\vec{\tau }\cdot \vec{\delta } \right )\right ]\psi _{B}  \nonumber \\
&+\frac{1}{2}\left (\partial ^{\mu } \sigma  \partial _{\mu }\sigma - m_{\sigma }^{2}\sigma ^{2} \right ) - \frac{\lambda _{1}}{3}\sigma  ^{3}-\frac{\lambda _{2}}{4}\sigma  ^{4} -\frac{1}{4}\Omega ^{\mu \nu }\Omega _{\mu \nu } +\frac{1}{2}m_{\omega }^{2}\omega _{\mu }\omega ^{\mu } +\frac{\lambda_3}{4}\left (\omega _{\mu }\omega ^{\mu }  \right )^{2} \nonumber \\
&-\frac{1}{4}\vec{P}^{\mu \nu }\cdot \vec{P}_{\mu \nu }+\frac{1}{2}m_{\rho }^{2}\vec{\rho }_{\mu }\cdot \vec{\rho }\,^{\mu }+\frac{1}{2}\left ( \partial ^{\mu }\vec{\delta }\cdot\partial _{\mu }\vec{\delta }-m_{\delta }^{2}\vec{\delta }\, ^{2}\right ) +\frac{\alpha _{3}{}'}{2}\,\omega _{\mu }\omega ^{\mu }\vec{\rho }_{\mu }\cdot\vec{\rho }\,^{\mu } \nonumber \\
&+g_{\sigma B}g_{\omega B}^{2}\sigma \omega _{\mu }\omega ^{\mu }\left ( \alpha _{1}+\frac{\alpha _{1}{}'}{2}g_{\sigma B}\sigma  \right )
+g_{\sigma B}g_{\rho B}^{2}\sigma \vec{\rho }_{\mu }\cdot\vec{\rho }\,^{\mu }\left ( \alpha _{2}
+\frac{\alpha _{2}{}'}{2}g_{\sigma B}\sigma  \right ) 
, \label{nlwm}  
\end{eqnarray}
with the index ${B}$ extending over the baryons. The meson mass is denoted by $m_i$, with $i=\sigma,\omega,\rho,\delta$, and $g_{iB}$ stands for the coupling constant of the interaction of the $i$ meson field with the baryonic field $\psi_B$. Parameters $\lambda_j$ and $\alpha _{k}$ are respectively related to self-interactions and  to  cross-interactions between mesonic fields. The field antisymmetric tensors read $\Omega _{\mu \nu}=\partial_\nu\omega_\mu-\partial_\mu\omega_\nu$ and $\vec{P} _{\mu \nu }=\partial_\nu\vec{\rho }_\mu-\partial_\mu\vec{\rho }_\nu-g_{\rho B}\left(\vec{\rho }_\mu \times \vec{\rho }_\nu \right)$. Yet, $\gamma^{\mu }$ and $\vec{\tau }$ are the Dirac gamma matrices and Pauli matrices for the isospin, respectively.

Due to the extremely high energy densities found in compact star cores, the existence of more massive baryonic species is expected \cite{hyp}. Those of the baryonic octet are considered next, so that $B=\{N,H\}$, where $N=\{p,n\}$ and $H=\{\Lambda^0, \Sigma^+, \Sigma^0, \Sigma^-, \Xi^0, \Xi^-\}$. One can define $g_{iB}=\chi_{iB}g_i$, where the set of constraints of the hyperon coupling scheme are taken as $\chi _{\sigma H}=\chi _{\delta H}=0.7$ and $\chi _{\omega H}=\chi _{\rho H}=0.783$, with $\chi _{iN}\equiv 1$ by construction \cite{Glen,hcs}.  Other choices, either based on quark coupling and SU(3) symmetry group \cite{Glen,lll} or  phenomenological adjustment of potential depths \cite{james1,james2} are also possible, 
but to avoid including extra uncertainties in our calculations, we have opted for the more common choice in the literature.

The absence of a hegemonic model for dense matter implies that dense hadronic matter can be described in different ways. RMF models based on the Lagrangian density given in eq. (\ref {nlwm}) depend on 16 free parameters to be determined somehow. In \cite {const1, const2} 263 choices presented in the literature were analyzed, confronting their predictions with well-established experimental results for nuclear matter, e.g., the compressibility modulus, the symmetry energy and its slope taken at the nuclear saturation point ($\rho_0$), and with the observational recent discovery of two massive pulsars with masses of the order of $2 ~M_ \odot$ \cite{2m1,2m2}.  According to ref. \cite {const1} only 35 NLWM parameterizations were approved by nuclear matter empirical constraints, all of them contained in two main categories: (i) models in which the couplings are density dependent and include the $\delta$ meson, and (ii) models that consider mesonic field cross-interaction terms. In the present study, we only consider type (ii) models. 
By requiring the  pre-approved RMF models to be able to describe stars with masses of the order of $2 ~ M_ \odot$,  only two of them are not discarded by the observational criteria when hyperons are included in the calculations, both of them of the category (i). The natural inclusion of hyperons in the description of matter softens the EoS and, as a consequence, decreases the maximum gravitational mass sustained by matter. 
The difficulty in reconciling the high measured masses of neutron stars with the description of these objects when there is the presence of hyperons in their interior is called {\it hyperon puzzle}, and its solution is one of the hottest research topics of nuclear astrophysics recently \cite {hyp}. On the other hand, in ref. \cite{const2} it was shown that models with cross-interaction on the mesonic fields are capable of producing maximum masses in the range of $1.93\le M/M_\odot \le 2.05$, at least when the effect of the hyperons are disregarded, and there are 11 parameterizations in this situation. The selected parameterizations are included in this class, and are the \textsc {IU--FSU} \cite {iufsu}  and the  \textsc{NL3$\omega\rho$} \cite {nl3wr} parameterizations.

\subsection{Deconfined quark phase}

The MIT bag model \cite{mit} has been widely used to describe quark matter, either confined in the hadrons substructure or unconfined in the form of QGP. It is a simple phenomenological model, whose Lagrangian reproduces the dynamics of the quark fields $\psi_{q}$ contained in a colorless region with volume $V$ delimited by the surface $S$ \cite{greiner1},
\begin{equation}
  \mathcal{L}_{\textsc{MIT}}= \sum_{q}\left [\bar{\psi }_{q}\left ( i\gamma ^{\mu} \partial _{\mu }-m_{q}\right)\psi_{q}-B  \right ]\Theta_{V} -\frac{1}{2}\bar{\psi }_{q}\psi_{q}\delta_{S},
 \label{eq:mitl}
\end{equation}
with the index ${q}$ extending over the quark flavor with mass $m_q$ with $q=u,d,s$. $\Theta_{V}$ is the Heaviside function, which guarantees the complete confinement of the wave functions of the quarks within the bag region, $\delta_{S}$ is a Dirac function, which ensures continuity of the fields on the surface $S$, and $B$ is the so-called bag constant, which represents a constant positive energy density needed to keep this region in the vacuum. 

Inside of the bag volume, the quarks are non-interacting and have kinetic energy, and no color currents go through the surface. Hence, if the energy at the border of the bag is negligible when compared with the energies inside it, the quarks in the bag interior can be taken as a Fermi gas. To the spherical bag of radius $R$, the $\Theta_{V}$ and $\delta_{S}$ argument turns $(R-r)$. So, the EoS can then be easily obtained from well-known thermodynamic results, and  the pressure and energy
density read
\begin{equation}
 P = \frac{1 }{ \pi^2} \sum_{q}\int_{0}^{{p_{F}}_q} dp\, \frac{p^4}{\sqrt{p^2+m_{q}^{2}}}-B
 \label{eq:pmit}
\end{equation}
and
\begin{equation}
 \varepsilon= \frac{3 }{ \pi^2}\sum_{q} \int_{0}^{{p_{F}}_q} dp\, \phantom{.} {p^2}{\sqrt{p^2+m_{q}^{2}}}+B,
  \label{eq:emit}
\end{equation}
with the Fermi level momentum written in terms of baryonic density ${p_{F}}_q= \sqrt[3]{3\pi^2 \rho_{q}}$.
The bag constant $B$ is a free parameter of the theory, but we next choose 
values that satisfy the Bodmer-Witten conjecture and the stability window according to ref. \cite{stab} , i.e.,
$$148  ~MeV \le B^{1/4} \le 168 ~MeV.$$

\subsection{Application to compact star description}

The models presented in previous sections were originally developed for application in the microscopic context, i.e. in nuclear matter, heavy ion collisions or in the hadron substructure analysis. Hence, some equilibrium conditions have to be imposed to the appropriate application of their EoS to the compact star description. Compact stars are considered as electrically neutral objects \cite{Glen}, so that stellar matter must consist of several particle species, negatively and positively charged. In this context, hadronic matter can be severely asymmetric in relation to isospin while nuclear matter is, in general, almost symmetrical. A non-interacting lepton gas is included in both descriptions in order to guarantee this equilibrium condition, so the charge neutrality implies for hadronic matter,
\begin{equation}
 \rho _{p}+\rho _{\Sigma ^{+}}=\rho _{\Sigma ^{-}}+\rho _{\Xi ^{-}}+\rho _{e^{-}}+\rho _{\mu ^{-}},
\end{equation}
and for strange matter,
\begin{equation}
 \frac{2}{3}\rho _{u}=\frac{1}{3}\rho _{d}+\frac{1}{3}\rho _{s}+\rho _{e^{-}}+\rho _{\mu ^{-}},
\end{equation}
where electrons and muons are the leptons considered.

The strangeness quantum number is not conserved in the compact star formation time scale, so a series of direct and inverse Urca processes can take place \cite{dec}. This decay reaction can be written generally as,
\begin{equation}
B_1\rightarrow B_2+\beta+\bar{\nu}_{\beta} \rightleftharpoons B_2+\beta\rightarrow B_1+{\nu}_{\beta} , \label{urca}\end{equation} 
where $ B_i $ can represent any baryons, since respecting energy and charge conservation, and $ \beta $ is a negatively charged lepton associated with the respective neutrino (anti neutrino) ${\nu}_{\beta}$ ($\bar{\nu}_{\beta}$). As the particles inside the compact stars are in the degenerated state, the matter will be in its state of equilibrium when the two reactions of the Urca process (\ref{urca}) reach equilibrium. This occurs when there are no more energy levels accessible to the leptons produced in direct $\beta$ decay. This chemical equilibrium condition turns to be, after the deleptonization phase in which neutrinos/anti neutrinos have already left the object carrying with it the thermal energy of the birth of the compact star, for the hadronic matter,
\begin{equation}
 \begin{gathered}
  \mu _{n}=\mu _{\Lambda ^{0}}=\mu _{\Sigma  ^{0}}=\mu _{\Xi ^{0}}, \\
  \mu _{p}=\mu _{\Sigma  ^{+}}=\mu _{n}-\mu _{e^-}, \\
  \mu _{\Sigma ^{-}}=\mu _{\Xi  ^{-}}=\mu _{n}+\mu _{e^-},
 \end{gathered}
\end{equation}
and for quark matter,
\begin{equation}
 \mu _{s}=\mu _{d}=\mu _{u}+\mu _{e^-},
\end{equation}
with $ \mu _{e^-}=\mu _{\mu^-}$ in both cases.

Given the EoS for hadronic and deconfined quark matter, respectively derived from the effective models (\ref{nlwm}) and (\ref{eq:mitl}), and taking into account the charge neutrality and chemical equilibrium conditions, the complete description of the compact star can be obtained from the solution of the Tolman-Oppenheimer-Volkoff (TOV) equations for relativistic hydrostatic equilibrium for the respective EoS \cite{tov1,tov2}. The main properties of the hyperonic and strange stars, as gravitational and baryonic masses, radii, and central energy densities and pressures can be then computed. The baryonic mass are of special relevance for this work, and can be written in terms of the energy density $\varepsilon$ and barionic number density $\rho$ as \cite{mtw}
\begin{equation}
 M_b=4\pi \int_0^R dr\, \left [ 1-\frac{2\, \varepsilon\! \left ( r \right )}{r} \right ]^{-1/2} r^2 \rho\! \left ( r \right ). \label{eq:massabar}
\end{equation}

\begin{figure}[t]
\centering

   \resizebox{0.8\linewidth}{!}{\input{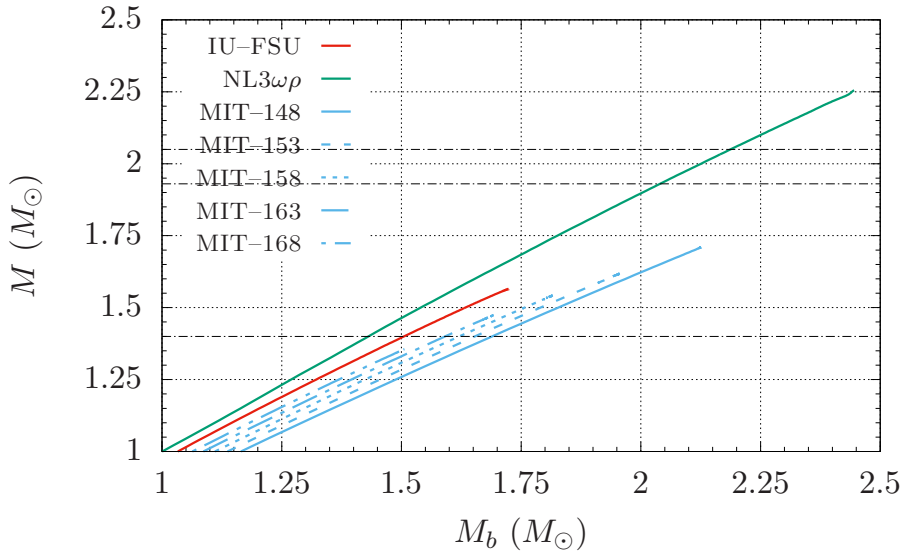}}
\caption{Gravitational versus baryonic mass for compact star families, obtained from the solution of the TOV equations with the EoS of the hyperonic and strange matter.}      \label{fig:mxm}
\end{figure}

In figure \ref{fig:mxm}, we show the gravitational versus baryonic masses obtained from the TOV equations to the EoS of hadronic and quark matter. The horizontal dashed lines represent the observational constraints, from top to bottom the first two delimit the band $ 1.93 \leq M / M_\odot \leq 2.05 $, which contains masses of the super-massive pulsars PSR J1614-2230 and PSR J0348+0432 \cite{2m1,2m2}, and the third denotes the Chandrasekhar limit $ M = 1.4 ~ M_\odot $ \cite{Glen}. Finally, table \ref {tab:nsprop} shows the main physical and observational characteristics of hadronic stars, obtained when considering the parameterizations discussed here conditioning the constituent hyperonic matter to charge neutrality and chemical equilibrium and including the BPS EoS for the description of the low-density matter in hadronic star crust \cite{BPS}.

\begin{table}[t]
\begin{center}
{  \begin{tabular}{c|cccc}
    \cline{2-5}
  & \multicolumn{2}{c}{\textsc{IU--FSU}} & \multicolumn{2}{c}{\textsc{NL3$\omega \rho$}} \\ 
  & ($p,n$) & ($8H$)  &  ($p,n$) & ($8H$) \\ \hline \hline
$M_\text{max}$ ($ M_\odot$)&$1.97$&$1.56$ & $2.76$ & $2.25$\\ \hline
$R$ ($km$)&$11.4$&$11.1$&$13.2$&$13.3$\\ \hline
{$P_c$}  ($MeV/fm^3$)&$350$&$235$&$454$&$206$ \\ \hline
{$\varepsilon_c$}  ($MeV/fm^3$)&$1216$&$1270$&$887$&$739$ \\ \hline
$R_{1.4}$ ($km$)  &$12.9$&$12.5$&$14.5$&$14.5$\\ \hline
  \end{tabular}}
   \caption{Main characteristics of nucleonic ($p,n$) and hyperonic ($8H$) stars with parameterizations \textsc {IU--FSU} and \textsc{NL3$\omega \rho$}. $ M_\text{max}$ is the maximum gravitational mass sustained by the model, $ R $ stands for the radius of this star, $ P_c $ and $ \varepsilon_c $ are respectively its central pressure and central energy density, and  $R_{1.4}$ denotes the radius of a star whose mass is that of the Chandrasekhar limit.}\label{tab:nsprop}
\end{center}
\end{table}

\section{Phase transition dynamics \label{sec3}}

The behavior of the graph shown in figure \ref {fig:mxm} suggests an interesting possibility. The conversion of a hadronic star into a strange star is energetically allowed, since the curve for hadronic stars is here always superior to that of the strange stars for a given baryonic mass, which agrees with the Bodmer-Witten hypothesis. We assume  the compact star as a pure hadronic star in the early stages after its emergence and first deleptonization, so the conversion of a hyperonic star into a strange star can take place. In this section we abord the phase transition process believed to occur in such cases, called quantum nucleation, summarizing its theory formalism.

\subsection{First order phase transitions}

The transition between the hadronic and the unconfined quark phases must occur in the strong interaction time scale, which is many orders of magnitude smaller than the weak interaction time scale. Consequently, the flavor must be conserved during the phase transition, which completely determines the composition of the quark phase of the final star from the hadronic matter in chemical equilibrium of the initial star through the bond
\begin{equation}
 y_q=\frac{1}{3}\sum_B n_{qB} y_B , \label{eq:flavour}
\end{equation}
where the baryonic number relative densities  $ y_i = \rho_i / \rho $ are related by the number $n_{qB}$ of $ q $ flavored quark constituents of baryon $ B $ \cite {nucleation}. In the case of static conversion processes, i.e., where there is no loss or accretion of matter, the total baryonic mass and the lepton number are also conserved, which consequently preserves the charge neutrality.
Thus, it is assumed that, at least under certain circumstances, the electrically neutral and in chemical equilibrium  hadronic matter (H-phase) is metastable and can be converted into an energetically favored, deconfined quark phase. Due to the imposition (\ref {eq:flavour}), this matter will not be in $ \beta $-equilibrium (hence, called Q*-phase). We next extend this notation ($^*$) to the EoS not in $\beta$-equilibrium. The chemical equilibrium will be readily reestablished by the quark matter through the Urca process, until it reaches the lowest energy state  in the form of the Q-phase.

{ In the present work, the deconfinement transition between the hadronic and QGP phase is described as a first-order phase transition,  obtained from the matching of two different models.}
The phase transition happens after the over-pressured metastable matter reaches the static transition point, defined according to the Gibbs criteria for the phase coexistence \cite{brach,bomb1}, 
\begin{equation}
\begin{gathered}
 T^{\left ( H \right )}=T^{\left ( Q^* \right )}=T,\\
  P^{\left ( H \right )}=P^{\left ( Q^* \right )}=P_0,\\
  \mu^{\left ( H \right )}(P_0,T)=\mu^{\left ( Q^* \right )}(P_0,T)=\mu_0,
\end{gathered} \label{eq:gibbscon}
\end{equation}
for the transition between phases $f=\{H,Q^*\}$ considered homogeneous, with
\begin{equation}
 \mu^{\left ( f \right )}=\frac{\varepsilon^{\left ( f \right )}+P^{\left ( f \right )}-s^{\left ( f \right )}T}{\rho^{\left ( f \right )}},
\end{equation}
where $\varepsilon^{\left ( f \right )}, P^{\left ( f \right )}$ e $\rho^{\left ( f \right )}$ are the total energy density, pressure and number density, deduced from the effective model, and $s^{\left ( f \right )}$ stands for the entropy density.

\begin{figure}[t]
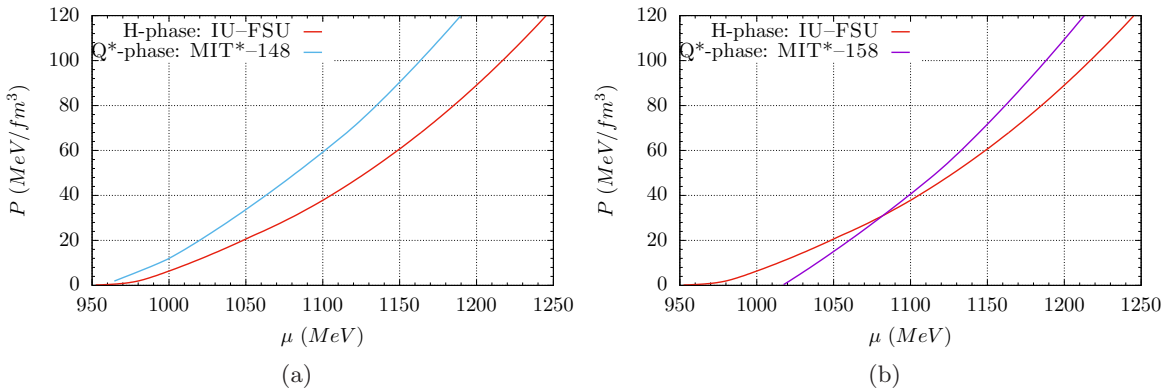

   \centering
   \subfloat[]{\label{fig:flava}\resizebox{0.5\linewidth}{!}{\input{flavmit1.tex}}}
      \subfloat[]{\label{fig:flavb}\resizebox{0.5\linewidth}{!}{\input{flavmit2.tex}}}
      \caption{Relation between pressure and chemical potential for the hadron and deconfined quark phases, respectively described by the \textsc{IU--FSU} parameterization and by the MIT* bag model.} \label{fig:flav}
\end{figure}

We consider here that the object is at $ T = 0 $ even during the phase transition process, which leaves only the values of $P_0$ and $\mu_0$ to be determined from the EoS of both phases. Even among the models considered suitable, the condition of coexistence of phases $ \mu ^ {\left ( f \right )} = \mu ^ {\left (Q ^ \ast \right)}$ can be satisfied or not, depending on the parameterizations used for the description of phases H and Q*. In figure \ref {fig:flav}  the procedure for checking and evaluating this phase coexistence condition is represented, showing in \ref {fig:flava} that the phase coexistence condition is not satisfied for $ B ^{1/4} = 148 ~ MeV $, which occurs for $ B ^{1/4} = 158 ~ MeV $, as seen in \ref {fig:flavb}. Table \ref {tab:gibbs} summarizes the results found for the other parameterizations considered in this work. Notice that the pressures are very low if the NL3$\omega\rho$ parametrization is used.

\begin{table}[t]
\begin{center}
  \begin{tabular}{c | cc}
    \cline{2-3}
{}&  \textsc{IU--FSU} & \textsc{NL3$\omega \rho$} \\ \hline \hline
\textsc{MIT*--148}  & No crossing & No crossing \\\hline
\multirow{2}{*}{{\textsc{MIT*--153}}}  &   $\mu_0=1022$ & $\mu_0=986$\\ 
& $P_0=12.4$  & $P_0=2.8$\\ \hline
\multirow{2}{*}{{\textsc{MIT*--158}}}  & $\mu_0=1082$&    $\mu_0=1016$\\ 
& $P_0=31.0$ &  $P_0=7.4$ \\ \hline
\multirow{2}{*}{{\textsc{MIT*--163}}}  & $\mu_0=1148$&  $\mu_0=1042$\\ 
& $P_0=60.0$ & $P_0=12.6$ \\ \hline
\multirow{2}{*}{{\textsc{MIT*--168}}}  & $\mu_0=1212$&   $\mu_0=1069$\\ 
& $P_0=97.3$ &  $P_0=19.1$\\ \hline
  \end{tabular}
   \caption{Values for $\mu_0$ (in MeV) and $P_0$ (in $MeV/fm^3$) for which the conditions of phase coexistence (\ref{eq:gibbscon}) are satisfied at $T=0$.}\label{tab:gibbs} 
\end{center}
\end{table}

\subsection{Lifshitz-Kagan theory}

\begin{figure}[t]
   \centering
   \resizebox{0.8\linewidth}{!}{\input{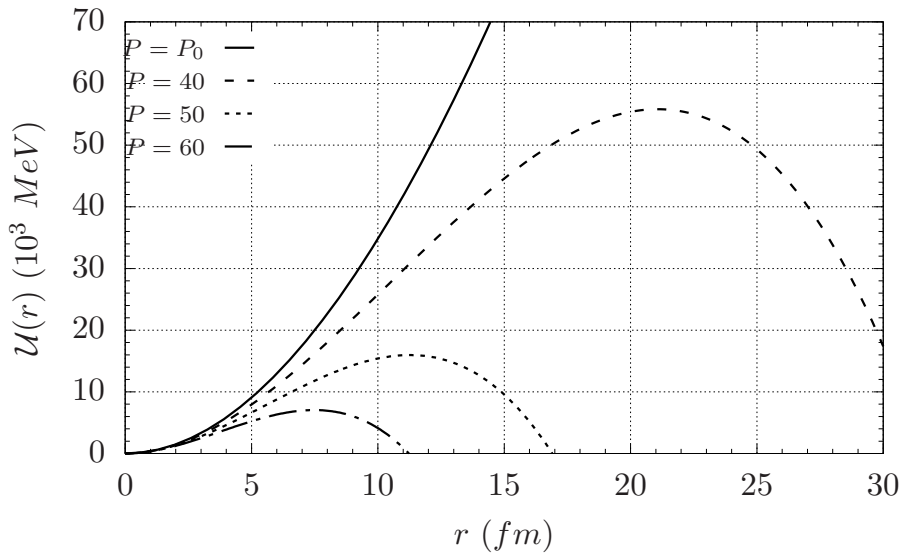}} 
   \caption{Potential energy barrier for the formation of a deconfined quark matter bubble in the metastable hadronic matter, for several values of pressure (in $MeV/fm^{3}$).}   \label{fig:u}
\end{figure}

A thermodynamic system is said to be metastable if its free energy is at a local minimum value, so that it remains stable for small fluctuations of the parameters, but remaining separated from the true ground state by a potential barrier in the configuration space. It allows the system to remain in this energetic unfavorable state for a long period of time \cite {ptd}. The emergence of a new thermodynamic phase in a system in metastable equilibrium through a first-order phase transition occurs through the process of nucleation, where the formation of a stable matter bubble in the metastable phase come from spatially localized fluctuations in the metastable matter, which lead to the emergence of regions of matter in the energetically favored phase and, under certain criteria, to the subsequent conversion of the whole system to this phase. The energy cost for the formation of a bubble of radius $r$ can be expressed in the form of a potential barrier \cite{nucleation,iida}, 
\begin{equation}
 \mathcal{U}(r)=\frac{4}{3}\pi \rho_f\left ( \mu _f - \mu_i\right )r^{3}+4\pi \sigma r^{2}, \label{eq:udrop}
\end{equation}
where the index $i$ and $f$ denote respectively the initial (hadronic) and final (quark) phases.
The terms of this equation are associated respectively with the volume and surface of the bubble, and relativistic corrections associated with the energy of curvature have been disregarded. 
{The surface tension constant $ \sigma $ is related to the amount of energy required to maintain the interface between hadronic and deconfined quark phases, which, as a consequence, influences the nucleation process quantitative description. This is a very important parameter because the allowance (or not) of the phase transition relies largely on its value, although it is not explicitly discussed in many cases, and its exact magnitude  remains rather undetermined.   There are several estimates of this value in the literature, as for instance, the ones obtained in \cite{dropl} and \cite{fragadrop}, based on the small surface thickness approximation, the former considering  hadronic matter while the second one  calculated for quark matter and predicting values of the order of  $5$--$15 ~ MeV / fm ^ 2$.  Another prescription, proposed in \cite{marcus}, was already tested in different calculations involving the construction of the pasta phase \cite{enjl,pastag}. In these calculations, both homogeneous and inhomogeneous phases were obtained with the same hadronic models, always remaining a free parameter to be evaluated. To avoid complications that would not contribute to our main discussion, in the present work we have opted to simply use the most accepted values.
It is usual in the literature to consider $\sigma$ in the range of $ 10$--$ 50 ~ MeV / fm ^ 2 $ \cite{sigmanuc}, and here we adopt these values.} 
In figure \ref {fig:u} we illustrate the behavior of the potential barrier (\ref{eq:udrop}) for the formation of a bubble of quark matter in the metastable hadronic matter, considering the H-phase modeled by parameterization \textsc{IU--FSU}, the Q*-phase described by the MIT* bag model for $B^{1/4}=158 ~MeV$, and taking $\sigma=30 ~MeV/fm^2$. 

Spontaneous processes lead to the reduction of the system free energy, so the stability of the drop depends on their radius in relation to the critical radius $ r_c $ where the energy barrier has a maximum, explicitly $r_c={2\sigma}/{\rho_f \left ( \mu _i - \mu_f\right )}$.
The activation energy required to form a stable matter bubble in the metastable phase with pressure $P>P_0$ may be provided by thermal fluctuations, shock waves, impurities in the constituent material or any other process which generates variations in the local properties of the system. Statistical oscillations of thermal origin must be dominant in the processes of finite temperature nucleation, but in the context of this work, where $ T = 0 $, the formation of this first critical nucleus occurs exclusively due to quantum effects, especially quantum tunneling \cite{nucleation,iida,onuki}. This process was first investigated by Lifshitz and Kagan \cite{lifkag} and provides the basic tools for the quantitative description of the phenomenon, especially in obtaining the  bubble formation rate and the metastable phase half-life.

Taking the approximation of the bubble growth rate $ \dot {r}$ sufficiently smaller than the velocity of sound in the medium, both phases can be considered incompressible. It is also assumed that the system adjusts adiabatically to the variations of $ r $, i.e., there is no dissipation and the process is reversible, so that the Lagrangian describing the growth of a spherical bubble of radius $ r $ can be written as
\begin{equation}
 L\!\left (r,\dot{r}\right)=-\mathcal{M}(r)c^2\sqrt{1-\left(\frac{\dot{r}}{c}\right)^{2}}+\mathcal{M}(r)c^2-\mathcal{U}(r), \label{eq:droplagr}
\end{equation}
where $\mathcal{U}(r)$ is the potential energy given in eq. (\ref{eq:udrop}) and
\begin{equation}
\mathcal{M}(r)=4\pi \varepsilon _{i}\left ( 1-\frac{\rho _{f}}{\rho _{i}} \right )^{2}r^{3} \label{eq:dropmass}
\end{equation}
is the bubble effective mass. Relativistic effects must be taken into account because the potential barrier height $\mathcal{U}_0=\mathcal{U}\!\left(r_c\right)$ is of the order of the critical bubble rest energy $\mathcal{M}\!\left(r_c\right)c^2$.

The Hamilton-Jacobi equation resulting from (\ref{eq:droplagr}) reads
\begin{equation}
 \left ( \frac{\partial S}{\partial r} \right )^{2}c^2-\left ( \frac{\partial S}{\partial t}+\mathcal{U}-\mathcal {M}c^2 \right )^{2}-\left(\mathcal {M}c^2\right)^{2}=0, \label{eq:hje}
\end{equation}
where $S(r,t)$ is the action associated with the Lagrangian. Through an eikonal approximation, eq. (\ref{eq:hje}) can be rewritten as a Schr\"odinger-type equation
\begin{equation}
 \left [ -\hbar^2c^2\frac{\partial ^{2}}{\partial r^{2}} +\left ( \mathcal {U}-E \right )\left ( 2\mathcal {M }c^2+E-\mathcal {U}\right )\right ]\psi=0, \label{eq:schroddrop}
\end{equation}
which can be straightforwardly evaluated by the Wentzel-Kramers-Brillouin (WKB) semi-classical approach \cite{iida}.
Hence, a set of equations for the determination of the rate of bubble formation and the metastable phase half-life is derived from this formalism, which constitute the relativistic version of the Lifshitz-Kagan theory, summarized next.

The fundamental state energy of the bubble, bound around $ R = 0 $, is denoted as $ E_0 $ and can be determined by the Bohr quantization rule
\begin{equation}
I\!\left ( E_0 \right )=2\pi \left ( m_0+\frac{3}{4} \right )\hbar, \label{eq:bohr}
\end{equation}
where $I\!\left ( E\right )$ is the action for the zero-point oscillation
\begin{equation}
 I\!\left ( E\right )=\frac{2}{c}\int_{0}^{r_{-}}\!dr\,\sqrt{\left ( 2\mathcal{M}c^2+E-\mathcal{U}\right )\left (E-\mathcal{U}\right )}, \label{eq:apz}
\end{equation}
and $m_0$ is the integer defined as
\begin{equation}
 m_0=\left \lfloor\frac{ I\!\left ( E_{min}\right ) }{2\pi\hbar}+\frac{1}{4}\right \rfloor,
\end{equation}
with $\left \lfloor {~} \right \rfloor$ denoting the floor function and $E_{min}$ standing for the smallest value of $E$ where the condition
$ 2\mathcal{M}c^2+E-\mathcal{U}\geq0$
is satisfied for arbitrary $r$, i.e., the lower limit for the energy values in which positive energy states occur. From this, the oscillation frequency $ \nu_0 $ can be set for the bubble interface as
\begin{equation}
 \nu _{0}^{-1}=\frac{\partial I}{\partial E},
\end{equation}
taken at $E=E_0$. 
Similarly, the probability $ p_0 $ that penetration occurs in the barrier follows from the action under the potential, namely, 
\begin{equation}
 A\!\left ( E\right )=\frac{2}{c}\int_{r_{-}}^{r_{+}}\!dr\,\sqrt{\left ( 2\mathcal{M}c^2+E-\mathcal{U}\right )\left (\mathcal{U}-E\right )}, \label{eq:asp}
\end{equation}
and reads
\begin{equation}
 p_{0}=\exp\left [ -\frac{A\!\left ( E_0 \right )}{\hbar } \right ]. \label{eq:prob}
\end{equation}
In the expressions above for $ I $ and $ A $, $ r_-$ ($ r_ +$) represents the smallest (largest) classical return points which delimit the classically forbidden region under the potential  barrier \cite{bomb1,iida}.

\begin{figure}[t]
   \centering
   \resizebox{0.8\linewidth}{!}{\input{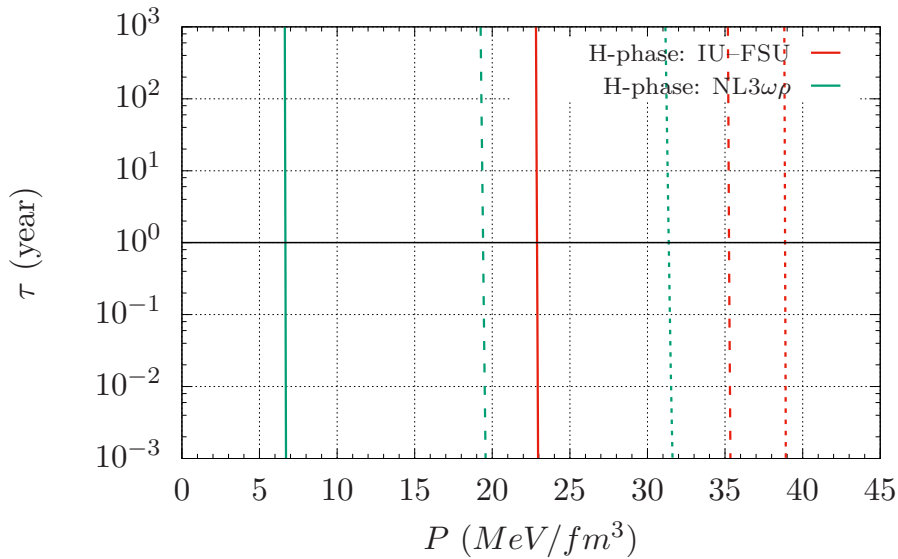}}
      \caption{Nucleation time $ \tau $ as a function of pressure $ P $ in the phase transition, considering the Q*-phase described by \textsc{MIT*--153}. The solid, dashed and dotted curves represent the values of $10$, $30$ and $50 ~MeV/fm^2$ for the parameter $\sigma$, respectively, and the horizontal line represents the nucleation time of $ \tau = 1$ year.}      \label{fig:tau}
\end{figure}

From equations (\ref{eq:bohr})--(\ref{eq:prob}), one can finally calculate the nucleation time $ \tau $ as
\begin{equation}
 \tau =\left ( N_{c}\nu _{0}p_{0} \right )^{-1}, \label{eq:tau}
\end{equation}
where $ N_c $ is the number of bubble forming centers. It is considered that the nucleation process occurs in the central part of the metastable neutron star ($ R \leq 100 ~m$), where the values for the EoS variables can be considered constant and equal to its central values in the star. Although there is uncertainty of one or two orders of magnitude in its value, we employ $ N_c = 10^{48}$ in agreement with the consulted bibliography \cite{iida,bomb1}.
In figure \ref {fig:tau} we show the behavior of the time required for the nucleation of the first bubble of matter in the final stable phase in terms of the pressure of the system, for the case of hadronic initial phase modeled by \textsc{IU--FSU} and \textsc {NL3$\omega \rho$} parameterizations and the final quark phase modeled by the \textsc{MIT*--153} model, as obtained through the Lifshitz-Kagan formalism. As one can see, $\tau$ changes abruptly with small variations in the pressure of the system during the phase transition. Our results indicate that $\tau$ can vary from many orders of magnitude larger than the universe age to almost zero in less than one percent increase of $P$.

\subsection{Astrophysical implications}

In the context of compact stars, nucleation was proposed as a relevant phenomenon  \cite {nucleation,nuchist} in order to clarify under which circumstances the transition between hadronic and deconfined quark phases would occur in the final evolution of massive stars ($M>8 ~M_\odot$), given Bodmer-Witten hypothesis of the stability of strange matter. The initial proposal was that this phase transition would occur during supernovae explosions, so that the released energy would even be  responsible for the final push that triggers the explosion process, which seems to be missing in the computational simulations of these events \cite {phaseboom}. Another possibility is that the formation of strange stars happens due to the {\it slow burning} of matter in a metastable hadronic star, triggered by nucleation, until the whole confined phase is converted into strange matter. Theoretically, both proposals are not mutually excludent, but only the second process is considered in this paper.

Analyzing eq. (\ref{eq:udrop}) for the potential barrier together with the behavior of the EoS for the H and Q* phases, we can see the tunneling which makes possible the appearance of stable matter demands an overpressure in relation to the coexistence pressure of the phases $ P_0 $,
\begin{equation}
 \Delta P=P-P_0>0.
\end{equation}
The greater this overpressure, the less opaque the barrier to be tunneled and the easier the nucleation of the first Q*-phase bubble in the metastable matter. Following ref. \cite{bomb1}, it is convenient to define the critical pressure $ P_{cr}$ as the pressure at which the half-life of the metastable phase is $ \tau = 1$ year. To this pressure is associated a critical gravitational mass $ M_{cr}$ of a hadronic star whose central pressure is $P_c=P_{cr}$. Since the star central pressure is directly related to its gravitational mass, stars with $ M>M_{cr}$ will strongly foment the occurrence of nucleation, thus producing a metastable phase half-life $ \tau <1$ year. We then take $ M_{cr}$ as a {\it maximum effective mass}  for hadronic stars, implying that hadronic stars with $ M>M_{cr}$ are very unlikely to be observed and replacing the Tolman-Oppenheimer-Volkoff limit, noting that while the TOV limit depends only on the hardness of the state equations of the hadronic matter, this new limit also depends on the EoS adopted for  the  deconfined quark matter and the interface parameter $ \sigma $. 

\begin{table}[t]
\begin{center}
 \subfloat[]{ \begin{tabular}{cc|cccc}
  \hline
$B^{1/4}$ & $\sigma$ & $P_{cr}$ & $\varepsilon$ & $M_{cr}$ & $R$  \\ \hline \hline
\multirow{3}{*}{$153$}&$10$&$22.8$&$296$&$0.9$&$12.9$\\ \cline{2-6}
& $30$&$35.2$&$381$&$1.1$&$12.8$\\ \cline{2-6}
& $50$&$38.8$&$403$&$1.2$&$12.8$\\ \hline
\multirow{3}{*}{$158$}&$10$&$42.0$&$421$&$1.2$&$12.8$\\ \cline{2-6}
& $30$&$55.7$&$491$&$1.3$&$12.6$\\ \cline{2-6}
& $50$&$59.0$&$507$&$1.3$&$12.6$\\ \hline
\multirow{3}{*}{$163$}&$10$&$70.5$&$565$&$1.4$&$12.5$\\ \cline{2-6}
& $30$&$81.3$&$618$&$1.4$&$12.3$\\ \cline{2-6}
& $50$&$85.1$&$636$&$1.4$&$12.3$\\ \hline
\multirow{3}{*}{$168$}&$10$&$109.8$&$746$&$1.4$&$12.0$\\ \cline{2-6}
& $30$&$122.1$&$797$&$1.5$&$11.9$\\ \cline{2-6}
& $50$&$126.6$&$816$&$1.5$&$11.9$\\ \hline
  \end{tabular}}\qquad
  \subfloat[]{  \begin{tabular}{cc|cccc}
    \hline
$B^{1/4}$ & $\sigma$ & $P_{cr}$ & $\varepsilon$ & $M_{cr}$ & $R$  \\ \hline \hline
\multirow{3}{*}{$153$}&$10$&$6.6$&$166$&$0.5$&$14.4$\\ \cline{2-6}
& $30$&$19.4$&$236$&$1.1$&$14.4$\\ \cline{2-6}
& $50$&$31.3$&$285$&$1.4$&$14.5$\\ \hline
\multirow{3}{*}{$158$}&$10$&$12.1$&$203$&$0.8$&$14.4$\\ \cline{2-6}
& $30$&$24.4$&$255$&$1.2$&$14.4$\\ \cline{2-6}
& $50$&$35.8$&$303$&$1.4$&$14.5$\\ \hline
\multirow{3}{*}{$163$}&$10$&$17.4$&$299$&$1.0$&$14.4$\\ \cline{2-6}
& $30$&$29.2$&$276$&$1.3$&$14.5$\\ \cline{2-6}
& $50$&$41.4$&$323$&$1.5$&$14.5$\\ \hline
\multirow{3}{*}{$168$}&$10$&$24.3$&$254$&$1.2$&$14.4$\\ \cline{2-6}
& $30$&$35.1$&$300$&$1.4$&$14.5$\\ \cline{2-6}
& $50$&$47.5$&$342$&$1.6$&$14.4$\\ \hline
  \end{tabular}
  }
   \caption{Main characteristics of hyperonic stars at critical nucleation conditions ($P_c=P_{cr}$), which result in a metastable H-phase half-life of $\tau=1$ year obtained with the parameterization (a) \textsc{IU--FSU} and (b) \textsc{NL3$\omega \rho$}. Here $B^{1/4}$ are in $MeV$ and other units are the same adopted previously in  text.}\label{tab:decay}
\end{center}
\end{table}

Hence, crossing the results of the Lifshitz-Kagan theory with the solutions of the Tolman-Oppenheimer-Volkoff equation it is possible to determine the characteristics of the hadronic stars that have $ P_c = P_ {cr} $ and that therefore, would be metastable and at the eminence of decaying into a strange star through a first order phase transition. As already discussed, this critical pressure depends not only on the modeling of the hadronic phase, but also on the parameters of the nucleation theory and the deconfined phase model. These results are summarized in table \ref {tab:decay} for the various combinations considered in this work.

\begin{figure}[t]
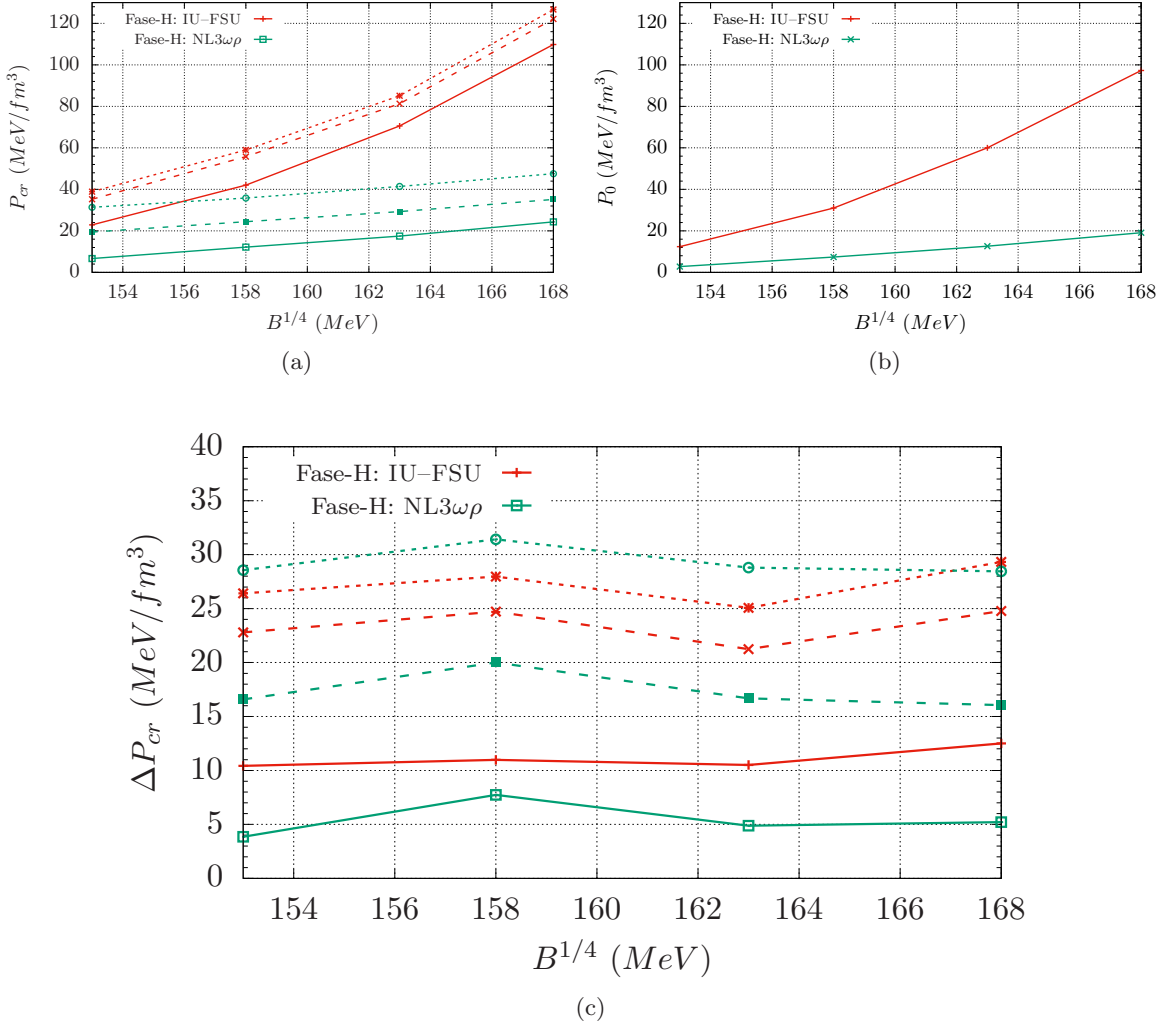

   \centering
   \subfloat[]{\label{fig:bagpa}\resizebox{.5\linewidth}{!}{\input{bagp.tex}}}
 \subfloat[]{\label{fig:bagpzero}\resizebox{.5\linewidth}{!}{\input{pzero.tex}}}
 \\
      \subfloat[]{\label{fig:bagpb}\resizebox{.8\linewidth}{!}{\input{bagpextra.tex}}}
         \caption{Relation between (a) the bag constant $B^{1/4}$ taken for the Q*-phase description and the critical pressure for the nucleation $P_{cr}$, (b) $B^{1/4}$ and the coexistence pressure of H and Q* phases $P_0$, and (c) $B^{1/4}$ and the overpressure $\Delta P_{cr}$. The solid, dashed and dotted curves in (a) and (c) represent the values of $10$, $30$ and $50 ~MeV/fm^2$ for the parameter $\sigma$, respectively.} \label{fig:bagp}
\end{figure}

In figure \ref{fig:bagpa} we show the critical pressure for the nucleation $P_{cr}$ as function of the bag constant adopted in the description of the final phase Q*. The results are strongly parameter-sensitive, depending on the parameterizations of both phases and on the value adopted for the surface tension constant $ \sigma $, which is an independent parameter. As expected, the critical pressure increases with the value of the bag parameter, which compensates for the degenerate quark gas internal pressure. { From table \ref{tab:gibbs}, we can see that the coexistence pressure increases with the increase of the bag parameter, which forces the critical pressure to be even larger for the nucleation process to take place. By plotting these results in figure \ref{fig:bagpzero}, it can be visualized that the dependence of both the critical pressure $P_{cr}$ and the coexistence pressure $P_0$ with the bag constant shows a  similar behavior. This interesting result can be seen if the critical configuration overpressure $ \Delta P_{cr} = P_{cr}-P_0 $ is plotted against the bag constant, as in figure \ref{fig:bagpb}, where the overpressure necessary for the existence of the nucleation process practically does not depend on the bag value, considering the small fluctuations in the curves as a result of approximations during the computational numerical procedure. In other words, these figures suggest that the overpressure depends basically on the model for the hadronic matter and the constant $ \sigma $ despite the fact that the results extracted from the description of matter in phase Q* are included in the calculations of the nucleation through equations (\ref{eq:udrop}) and (\ref{eq:dropmass}).} Due the complexity of the algorithms used in obtaining the results it is difficult to quantitatively describe the causes of this constant resistance to nucleation, but we suppose that it comes from bonds imposed to Q*-phase EoS in relation to H-phase during the transition.  Further conceptual understanding of its causes is still demanded. This behavior is verified for both parametrization of the NLWM used in the description of the hadronic matter, so it would be desirable to verify if this behavior is a property of the phase transition in the chromodynamic regime by testing it in more credible effective models, e.g., those that take into account quark matter chirality. 

\begin{figure}[t]
   \centering
\resizebox{0.8\linewidth}{!}{\input{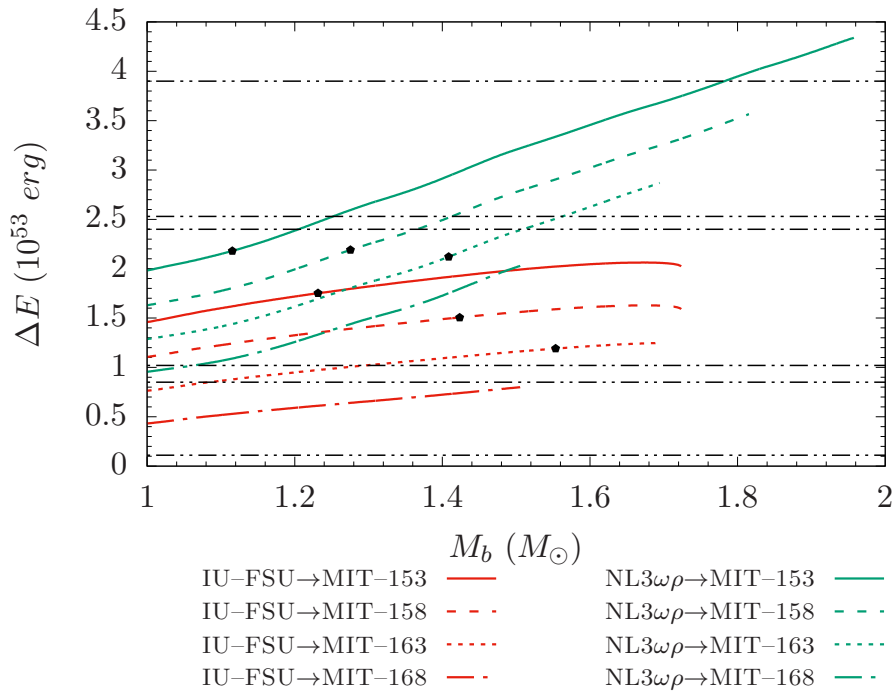}}
   \caption{Energy released by the total conversion of a hadronic star into a strange star. The dark points on the curves represent the star at the nucleation critical configuration when $\sigma=30 ~MeV/fm^2$. The horizontal dashed lines represent some observational values for GRB isotropic equivalent energies, see table \ref{tab:grb}.}\label{fig:edecay}
\end{figure}

Following refs. \cite{bomb1,bomb2} we assume that massive star remnants survive the early stages of their evolution, i.e., the supernova explosion and subsequent cooling by deleptonization, as a pure hadronic star and then converts to strange star through nucleation. 
Taking the baryonic mass conservation as the analogous of the baryonic number conservation during the conversion process, we can calculate the released energy in such decay processes from the well-known mass-energy relation, where the gravitational masses of the initial (constituted by H-phase matter) and final (constituted by Q-phase, after the $\beta$ equilibrium is reestablished) star are taken for the same baryonic mass \cite{debgrb,bombgrb}, explicitly given by
\begin{equation}
 \Delta E=\left(M_i-M_f\right)\times17.88\times10^{53}\text{ erg},
\end{equation} 
where $M_{k}$ stands for the mass of the $k=i,f$ star, in $M_\odot$ units. In figure \ref{fig:edecay}, the energy released by such conversion is shown. The dark points on the curves illustrate the star at the nucleation critical configuration for the case $\sigma=30 ~MeV/fm^2$. { For lower surface tensions, the nucleation process are favored, i.e., the critical configuration would have smaller masses than the marked in the figure, and it would have bigger masses for higher surface tensions. This point shows the minimum amount of energy released in the conversion for the considered parameter set, and the release of energy should happen only for masses  larger than that of the critical configuration. }

The huge amount of released energy can be  associated with neutrino bursts and gravitational wave emission, producing a second delayed explosion with respect to the first supernova explosion \cite{bomb2}.
As assumed in this work, metastable hadronic stars can be seen as GRBs progenitors, with the conversion process triggered by nucleation being the source of the huge amounts of energy of this events. Also, it suggests a delayed connection between supernova explosions and GRBs, the so-called quark deconfinement nova \cite{qdn}, and proposes an explanation to the observed bimodal distribution of the kick velocities of radio pulsars \cite{bimod}. The released energies obtained here by the application of effective models to the description of dense matter and by considering the Lifshitz-Kagan formalism, leads us to conclude that the considered hypothesis of the conversion of a hadronic star into a quark star as one of the possible causes of GRBs is reasonable, mainly for energies in the range $0.5<E_\text{iso}/10^{53}\text{ erg}<4.0$, as can be seen in figure \ref{fig:edecay} and table \ref{tab:grb}. This kind of process cannot provide, however, the amount of energy of more energetic GRB, e.g., the high redshift LGRBs 080607 ($z=3.036$) and 080721 ($z=2.591$) which released  $20.0(1.3)\times10^{53}\text{ erg}$ and $12.0(1.2)\times10^{53}\text{ erg}$ respectively \cite{grbdata1}. 
 Another mechanism that relates the conversion of a hadronic star into a quark star while inducing a short GRB \cite{angelez_2013} is
based on the internal self-annihilation of dark matter accreted from the galactic halo in the interior of the hadronic stars.
However, the large diversity of GRBs characteristics allow this class of phenomena and an assortment of progenitors not considered in the present work. 

\begin{table}[t]
\begin{center}
  \begin{tabular}{cc|cc}
    \hline
\multicolumn{2}{c|}{GRB  (type)} & $z$ & $E_\text{iso}$ ($\times10^{53}$ erg)   \\ \hline \hline
051221&(S)&0.5465&0.03(0.004)\\ \hline
070714&(S)&0.92&0.11(0.01)\\ \hline
080411&(L)&1.03&2.4(0.2)\\ \hline
080605&(L)&1.639&2.53(0.36)\\ \hline
071020&(S)&2.145&1.02(0.15)\\ \hline
080413&(L)&2.433&0.85(.10)\\ \hline
080810&(L)&3.35&3.9(0.37)\\ \hline
  \end{tabular}
   \caption{Some Long (L) and Short (S) gamma-ray bursts with measured redshifts $z$ and isotropic equivalent energies $E_\text{iso}$ according to ref. \cite{grbdata1}.} \label{tab:grb}
\end{center}
\end{table}

\begin{figure}[t]
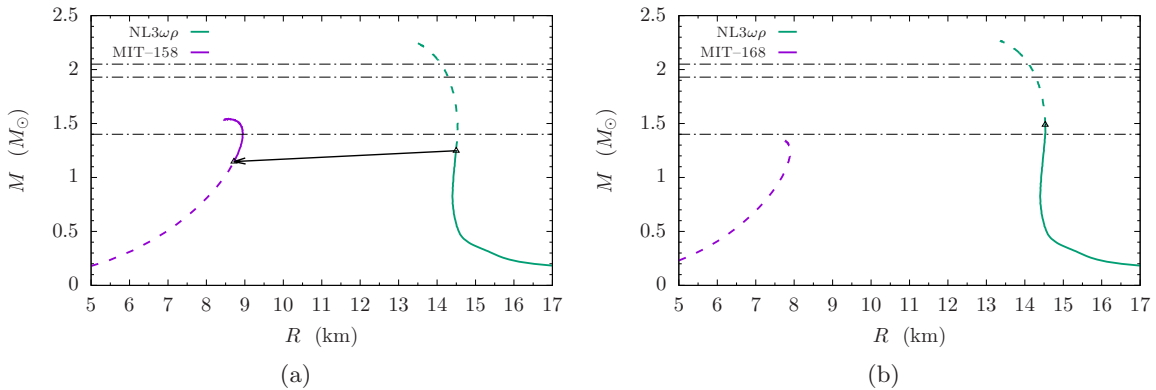

   \centering{
   \subfloat[]{\label{fig:conva}\resizebox{.5\linewidth}{!}{\input{conv.tex}}}
      \subfloat[]{\label{fig:convb}\resizebox{.5\linewidth}{!}{\input{conv2.tex}}}}
         \caption{Mass-radius relation for the hadronic star family obtained with the \textsc{NL3$\omega \rho$} parameterizations for the H-phase and for strange star family considering the bag parameter (a) $B^{1/4}=158 ~MeV$ and (b) $B^{1/4}=168 ~MeV$ on the modeling of the Q-phase, with$\sigma=30 ~MeV/fm^2$. The configurations marked with black dots are the stars at critical situation for nucleation (see table \ref{tab:decay}).}\label{fig:conv}
\end{figure}

According to the scenario proposed in this study for stellar final evolution, the existence of completely stable hadronic compact stars with masses exceeding a very restrictive threshold is unfavored, due the instability of the H-phase triggered by the nucleation process. This limit for hadronic star mass is strongly dependent on the set of parameters adopted on the calculations, being in the range of $0.9<M/M_\odot<1.5$ for the \textsc{IU--FSU} parameterization of the H-phase and in the range of $0.5<M/M_\odot<1.6$ for the \textsc{NL3$\omega \rho$}, allowing masses slightly bigger than the Chandrasekhar limit in both cases. Figure \ref{fig:conv} shows two interesting situations of the mass-radius curve for hadronic and strange stars families sustained by \textsc{NL3$\omega \rho$} EoS for the H-phase and considering $B^{1/4}=158 ~MeV$ (\ref{fig:conva}) and $B^{1/4}=168 ~MeV$ (\ref{fig:convb}) on the modeling of the Q-phase, taking $\sigma=30 ~MeV/fm^2$, and the horizontal dashed lines represent the theoretical and observational constraints (see figure \ref{fig:mxm}). The dashed branches represent the unstable configurations of hadronic stars and the strange star inaccessible through the slow burning process considered here. From these typical behaviors, we can infer that the existence of small radii quark stars does not exclude the coexistence of hadronic stars with the same gravitational mass and a larger radii. If the conversion is allowed (\ref{fig:conva}), more massive quark stars can also be produced by long-term mass accretion onto the compact star after the decay. But, if the the nucleation in the hadronic star triggers the phase transition at masses greater than the sustained by the quark matter EoS (\ref{fig:convb}), the metastable hadronic star can not decay to a stable strange star and the conversion process leads to the formation of a black hole \cite{bomb1,bomb2}. Anyway, the model combinations considered in this work do not allow the description of $2 ~M_\odot $ compact stars under any circumstances when the deconfinement phase transition is taken into account.

\section{Final remarks and conclusions \label{sec4}}

In the present work, we considered the possibility of pulsars as hadronic and strange quark stars, both cases described in the framework of relativistic effective models for dense matter, respectively the NLWM and the MIT bag model.
Previous calculations suggested that the results depend  on the parameterizations considered for the models. 
To check how strong the model dependence really is, we have chosen parameterizations for the hadronic EoS which have been tested and approved with respect to nuclear matter properties and astrophysical observational constraints \cite{const1,const2} and also restricted the bag parameters of the deconfined quark matter model according to the stability window for which the Bodmer-Witten hypothesis is fulfilled \cite{stab}.

From the descriptions of compact stars given by the effective models, we analyzed the conditions in which the phase conversion between hadronic and strange stars is allowed. The determinant process in this situation is the so-called quantum nucleation, and its formalism was summarized. The results arising from the EoS adopted suggest that hadronic stars are metastable to the decay to strange stars or even black holes above a threshold value of the gravitational mass. The lifetime of the metastable star was calculated and used to redefine the concept of limiting mass of compact stars through the critical pressure of nucleation. While the TOV limit \cite{tov1,tov2} depends only on the hadronic matter EoS stiffness, the critical mass $M_{cr}$ is strongly parameter-sensitive on the parameterizations adopted for both phases description and on the value adopted for the surface tension constant $\sigma$. However, it can be seen that the overpressure $ \Delta P_{cr} = P_{cr}-P_0 $ necessary for nucleation does not depend on the bag value adopted on the description of deconfined quark matter. This constant resistance to nucleation was not discussed in the literature before  { because it is was not obvious.  Previous results neither suggested that the coexistence pressure and the critical nucleation pressure show the same dependence on the bag parameter in such a way that $\Delta P_{cr}$ is roughly constant nor that, for a given hadronic EoS, the same overpressure above the coexistence pressure results in the same nucleation time independently of the deconfined quark matter EoS.}
To verify if this behavior is a  simple property of the chromodynamic phase transitions, the authors intend to apply the methods discussed above to more credible models in a forthcoming work. { One possible improvement in the description of quark matter is the use of an EoS that considers the interaction between quarks and gluons in a more rigorous way, as proposed in \cite{fragaq}. }

The conversion of a hadronic star into a strange star is energetically allowed by the results obtained from the effective models. The hypothesis of GRBs as manifestations of the energy released in the conversion of a metastable hadronic star into a strange star is also verified, as we obtained released energies in the range $0.5<E_\text{iso}/10^{53}\text{ erg}<4.0$, agreeing with the extensive analysis of ref. \cite{debgrb}.  When the deconfinement phase transition is taken into account, we could infer the coexistence of small radii quark stars and larger radii hadronic stars in a narrow band of gravitational masses, if the conversion is allowed by having an accessible stable strange star at the mass where the nucleation is triggered. Yet, the decay to a strange star is not viable depending on the model combination considered, leading instead to the formation of a black hole from metastable hadronic stars with masses greater than $M_{cr}$. Even considering that the decay is allowed and more massive quark stars can also be produced by late mass accretion, under any circumstances the model combinations considered in this work allow the description of $2 ~M_\odot $ compact stars. Therefore, a new astrophysical constraint for effective model parameterizations must be added to these considered in refs. \cite{const1,const2,stab} to demand the description of stable strange stars with masses above the observational bond if the decay is allowed, or with a $M_{cr}$  of the order of  $2 ~M_ \odot$ if it is not.

\acknowledgments
The authors thank Prof. Constan\c ca Provid\^encia for helpful discussions.
This work is a part of the project INCT-FNA Proc. No. 464898/2014-5,
was partially supported by CNPq (Brazil) under grant 300602/2009-0 and CAPES (Brazil).


\begin{thebibliography}{99}

\bibitem{bell} A. Hewish et al,
Nature, $\mathbf{217}$, 713 (1968).

\bibitem{greiner1} W. Greiner, S. Schramm, E. Stein, 
\emph{Quantum Chromodynamics}, 
Springer, New York - 2nd. ed. (2002).

 \bibitem{schmitt} A. Schmitt,
\emph{Dense Matter in Compact Stars},
Springer,  Berlin   (2010).

\bibitem{Glen} N.~K. ~Glendenning,
\emph{Compact Stars - Nuclear physics, Particle physics, and General relativity},
Springer, New York -  2nd. ed. (2000).

 \bibitem{astrolab} F. Weber et al,
Prog. Part. Nucl. Phys. $\mathbf{59}$, 94 (2007).

 \bibitem{pmass} B. Kiziltan et al,
Astrophys. J. $\mathbf{778}$ (2013).

\bibitem{ito} N. Itoh, Prog. Theor. Phys. $\mathbf{44}$, 291 (1970).

\bibitem{bw1} A. R. Bodmer,
Phys. Rev. D $\mathbf{4}$, 1601 (1971).

\bibitem{bw2} E. Witten,
Phys. Rev. D $\mathbf{30}$, 272 (1984).

\bibitem{bw3} E. Fahri, R. L. Jaffe,
Phys. Rev. D $\mathbf{30}$, 2379 (1984).

 \bibitem{qsweber} F. Weber,
Prog. Part. Nucl. Phys. $\mathbf{54}$, 193 (2005).

 \bibitem{qgp1} H. Bohr, H. B. Nielsen,
 Nucl. Phys. B $\mathbf{128}$, 275 (1977).
 
 \bibitem{qgp2} O. K. Kalashinov, V. V. Kimov,
 Phys. Lett. B $\mathbf{88}$, 328 (1979).

 \bibitem{qgpl} F. Karsch,
 Nucl. Phys. A $\mathbf{590}$, 367 (1995).
 
\bibitem{iv1} D. D. Ivanenko, D. F. Kurdgelaidze,
Astrophys. $\mathbf{1}$, 251 (1965).

\bibitem{iv2} D. D. Ivanenko, D. F. Kurdgelaidze,
Lett. Nuovo Cimento $\mathbf{2}$, 13 (1969).

\bibitem{qsexp1} I. Bombaci,
Phys. Rev. C $\mathbf{55}$, 1587 (1997).

\bibitem{qsexp2} R. X. Xu,
Astrophys. J. $\mathbf{570}$, L65 (2002).

\bibitem{qsexp3} J. J. Drake et al,
Astrophys. J. $\mathbf{572}$, 996 (2002).

\bibitem{grbdata1} G. Ghirlanda et al,
Astron. Astrophys. $\mathbf{496}$, 585 (2009).

\bibitem{grbdata2} T. Piran,
Rev. Mod. Phys. $\mathbf{76}$, 1143 (2004).

\bibitem{cgrb1} K. S. Cheng, Z. G. Dai,
Phys. Rev. Lett. $\mathbf{77}$, 1210 (1996).

\bibitem{cgrb2} I. Bombaci, B. Datta,
Astrophys. J. Lett. $\mathbf{530}$, L69 (2000).

\bibitem{debgrb} D. P. Menezes et al,
Phys. Rev. C $\mathbf{73}$, 025806 (2006).

\bibitem{const1} M. Dutra et al,
 Phys. Rev. C $\mathbf{90}$, 055203 (2014).

 \bibitem{const2} M. Dutra, O. Louren\c co, D. P. Menezes,
 Phys. Rev. C $\mathbf{93}$, 025806 (2016)
[Erratum: Phys. Rev. C $\mathbf{94}$, 049901 (2016)].

 \bibitem{urca} D. Page, J. H. Applegate,
Astrophys. J. Lett. $\mathbf{394}$, L17 (1992).

\bibitem{qhd1}  M. H. Johnson, E. Teller,
Phys. Rev. $\mathbf{98}$, 783 (1955).

\bibitem{qhd2}  H.-P. Duerr, E. Teller,
Phys. Rev. $\mathbf{101}$, 494 (1956).

\bibitem{walecka} J. D. Walecka,
\emph{Theoretical Nuclear and Subnuclear Physics},
World Scientific Publishing, London -  2nd. ed. (2004).

\bibitem{bogbod} J. Boguta, A. R. Bodmer, 
Nucl. Phys. A $\mathbf{292}$, 413 (1977).

\bibitem{nlwm1} B. D. Serot, J. D. Walecka, 
Adv. Nucl. Phys. $\mathbf{16}$, 1 (1986).

\bibitem{hyp} I. Vida\~na,
J. Phys.: Conf. Ser. $\mathbf{668}$ 12031 (2016).

\bibitem{hcs} D. P. Menezes, C. Provid\^encia,
Braz. J. Phys. $\mathbf{34}$, 724 (2004).

\bibitem{lll} L. L. Lopes, D. P. Menezes, 
Phys. Rev. C $\mathbf{89}$, 025805 (2014).

\bibitem{james1} J. R. Torres, F. Gulminelli, D. P. Menezes, 
Phys. Rev. C $\mathbf{93}$, 024306 (2016).

\bibitem{james2} J. R. Torres, F. Gulminelli, D. P. Menezes, 
Phys. Rev. C $\mathbf{95}$, 025201 (2017).

\bibitem{2m1} P. B. Demorest et al,
Nature, $\mathbf{476}$, 7319 (2010).

\bibitem{2m2} J. Antoniadis et al,
Science, $\mathbf{340}$ 6131 (2013).

\bibitem{iufsu} F. J. Fattoyev et al,
Phys. Rev. C $\mathbf{82}$, 55803 (2010).

\bibitem{nl3wr}  J. Fang et al,
Phys. Rev. C $\mathbf{95}$, 45802 (2017).

\bibitem{mit} A. Chodos et al, 
Phys. Rev. D $\mathbf{9}$, 3471 (1974).

\bibitem{stab} J. R. Torres, D. P. Menezes, 
Europhys. Lett. $\mathbf{101}$, 42003 (2013).

\bibitem{dec} R. R. Silbar, S. Reddy,
Am. J. Phys. $\mathbf{72}$, 892 (2004).

\bibitem{tov1} R. C. Tolman,
Phys. Rev. $\mathbf{55}$, 364 (1939).

\bibitem{tov2} J. R. Oppenheimer, G. M. Volkoff,
Phys. Rev. $\mathbf{55}$, 374 (1939).

\bibitem{mtw} C. W. Misner, K. S. Thorne, J. A. Wheeler,
\emph{Gravitation},
 W. H. Freeman \& Co., San Francisco (1973).
 
 \bibitem{BPS} G. Baym, C. Pethick, P. Sutherland,
 Astrophys. J. $\mathbf{170}$, 299 (1971).
  
 \bibitem{nucleation} M. L. Olesen, J. Madsen,
 Phys. Rev. D $\mathbf{49}$, 2698 (1994).

 \bibitem{brach} M. K. Brachman,
J. Chem. Phys.  $\mathbf{22}$, 115 (1954).

\bibitem{bomb1} I. Bombaci, I. Parenti, I. Vida\~na,
Astrophys. J. $\mathbf{614}$, 314 (2004).

\bibitem{ptd} W. S. Koch,
 \emph{Dynamics of first-order phase transitions in equilibrium and nonequilibrium systems},
 Springer-Verlag, Berlin (1984).
 
 \bibitem{iida} K. Iida, K. Sato,
 Phys. Rev. C $\mathbf{58}$, 2538 (1998).
 
 \bibitem{dropl} D. P. Menezes, C. Provid\^encia,
Nucl. Phys. A $\mathbf{650}$, 283 (1999).

\bibitem{fragadrop}  L. F. Palhares, E. S. Fraga
Phys. Rev. D  $\mathbf{82}$,  125018 (2010).

\bibitem{marcus} M. B. Pinto, V. Koch, J. Randrup,
Phys. Rev. C $\mathbf{86}$, 025203 (2012).

\bibitem{pastag} G. Grams et al,
Phys. Rev. C $\mathbf{95}$, 055807 (2017).

\bibitem{enjl} H. Pais, D. P. Menezes, C. Provid\^encia,
Phys. Rev. C $\mathbf{93}$, 065805 (2016).
 
 \bibitem{sigmanuc} H. Heiselberg, C. J. Pethick, E. F. Staubo, 
 Phys. Rev. Lett. $\mathbf{70}$, 1355 (1993).
 
 \bibitem{onuki} A. Onuki,
  \emph{Phase transition dynamics},
Cambridge University Press, Cambridge (2002).

\bibitem{lifkag} I. M. Lifshitz, Y. Kagan,
Zh. Eksp. Teor. Fiz. $\mathbf{62}$, 385 (1972) 
[Sov. Phys. JETP $\mathbf{35}$, 206 (1972)].

\bibitem{nuchist} G. Baym et al,
Phys. Lett. B $\mathbf{160}$, 181 (1985).

\bibitem{phaseboom} J. E. Horvath, O. G. Benvenuto, 
Phys. Lett. B $\mathbf{213}$, 516 (1988).

\bibitem{njl} Y. Nambu,  G. Jona-Lasinio,
Phys. Rev. $\mathbf{122}$, 345 (1961); $\mathbf{124}$, 246 (1961).

\bibitem{bombgrb} I. Bombaci, B. Datta,
Astrophys. J. Lett. $\mathbf{530}$, L69 (2000).

\bibitem{bomb2} I. Bombaci, D. Logoteta,
Int. J. Mod. Phys. D $\mathbf{26}$, 1730004 (2017).

\bibitem{qdn} Z. Berezhiani et al,
Astrophys. J. $\mathbf{586}$, 1250 (2003). 

\bibitem{bimod} Z. Arzoumanian, D. F. Chernoff, J. M. Cordes,
Astrophys. J. $\mathbf{568}$, 289 (2002).

\bibitem{angelez_2013} A. \'Angelez P\'erez-Garcia, F. Daigne, J. Silk,  Astrophys. J. $\mathbf{768}$, 145 (2013).

\bibitem{fragaq} E. S. Fraga, A. Kurkela, A. Vuorinen,
Astrophys. J. Lett. $\mathbf{781}$, L25 (2014).
\end{thebibliography}
\end{document}